\documentclass[runningheads]{llncs}

\usepackage{graphicx}
\usepackage{textcomp}
\usepackage[caption=false]{subfig}
\usepackage{amsfonts}
\usepackage[toc,page]{appendix}

\begin{document}
%
\title{Interpretation of 3D CNNs for Brain MRI Data Classification}
\titlerunning{Interpretation of 3D CNNs for Brain MRI}

%
\author{Maxim Kan\inst{1}\and
Ruslan Aliev\inst{1}\and
Anna Rudenko\inst{1}\and
Nikita Drobyshev\inst{1}\and
Nikita Petrashen \inst{1}\and
Ekaterina Kondrateva\inst{1}\and
Maxim Sharaev\inst{1}\and
Alexander Bernstein\inst{1}\and
Evgeny Burnaev\inst{1}}

\authorrunning{M. Kan et al.}

\institute{Skolkovo Institute of Science and Technology, Moscow, Russia, \\ 
\email{ekaterina.kondrateva@skoltech.ru}}
\maketitle              
\begin{abstract}


Deep learning shows high potential for many medical image analysis tasks. Neural networks can work with full-size data without extensive preprocessing and feature generation and, thus, information loss.
Recent work has shown that the morphological difference in specific brain regions can be found on MRI with the means of Convolution Neural Networks (CNN). However,  interpretation of the existing models is based on a region of interest and can not be extended to voxel-wise image interpretation on a whole image. In the current work, we consider the classification task on a large-scale open-source dataset of young healthy subjects --- an exploration of brain differences between men and women.
In this paper, we extend the previous findings in gender differences from diffusion-tensor imaging on T1 brain MRI scans. We provide the voxel-wise 3D CNN interpretation comparing the results of three interpretation methods: Meaningful Perturbations, Grad CAM and Guided Backpropagation, and contribute with the open-source library. 

\end{abstract}

\section{Introduction}

Deep learning and specifically Convolutional Neural Networks (CNNs) has recently found many applications in the area of medical diagnostics and image processing \cite{Sharaev2018} \cite{Pominova2019b}. For example, processing Magnetic Resonance Images (MRI) with CNN allow reducing the dose of gadolinium used for contrast by an order of magnitude \cite{gado}. Another example  is the detection of cerebral microbleeds using a 3D-CNN \cite{microbleeds}. 
In the following work \cite{segm} authors apply convolutional networks to multi-modal (T1, T2 and FA) MRI images in order to segment infant brain tissue images into Gray Matter (GM), White Matter (WM), and Cerebrospinal Fluid (CSF). CNNs are used in early-stage Alzheimer's disease detection in MRI and PET images \cite{alz}. Finally, these type of networks are applied to a variety of predictive regression tasks in brain imaging, see \cite{Pominova2019}. 

Conventionally, the brain data is firstly processed to get the lower dimensional meaningful features \cite{FADTI} or goes with whole-brain statistical analysis or Voxel-Based Morphometry (VBM). For the diffusion tensor imaging (DTI) extracted features conventionally include fraction anisotropy (FA),  mean, axial and radial diffusivity values \cite{main_article}. For functional T2* MRI images functional connectivity features are commonly extracted, and for T1 structural imaging --- morphometry features. Data analysis and machine learning usually follow this feature extraction step \cite{Sharaev2018_2}.

From the other side, deep learning approaches, especially those for processing 3D data, are shown to be more accurate in many applications \cite{main_article} as they use full-sized data without information loss during aforementioned extensive pre-processing and feature extraction.

However, the stable and reliable interpretation of 3D CNNs is still a big topic of discussion. For example, recent studies of age and gender brain differences point out that despite its high prediction quality ``the localised predictions of age and gender do not yield easily interpretable insights into the workings of the neural network'' \cite{pawlowski2019texture}.

Deep learning models interpretation in MRI implies training on large databases of healthy subjects. One of the most common and highly explored databases available in open-access is a Human Connection Project (HCP)\footnote{https://db.humanconnectome.org}. A conventional task being extensively explored within this database is a task of gender patterns recognition between men and women \cite{sex_matters}, \cite{evolving_sex}.

However, previous studies on morphological difference between specific brain regions show interpretable results only on the feature or region-of-interest level  \cite{main_article}. On the contrary, the state-of-the-art deep-learning interpretation methods allow visualization of the decision rule in a pixel-wise fashion. Or in the case of 3D convolution models --- voxel-wise \cite{Age_Sex_Mri}. The contributions of the proposed paper are as follows: 
\begin{itemize}
  \item we reproduce and extend the state-of-the-art 3D CNN model \cite{main_article} to investigate the difference between men and women brain on T1 images and confirm previous findings on DTI;
  \item we apply several network interpretation methods to the 3D CNN model: Meaningful Perturbations, Grad CAM and Guided Backpropagation to find gender-specific patterns and compare their performances;
  \item we compare the obtained attention maps to the conventional machine learning classification models on morphometry data and discuss the differences as well as with previous findings;
    \item lastly we provide the code for MRI 3D CNN interpretation as an open-source library.
\end{itemize}

The source code is open and available at \\ \texttt{https://github.com/maxs-kan/InterpretableNeuroDL}

\section{Data}

The database Human Connectome Project (HCP) contains MRI data from 1113 subjects, including 507 men and 606 women of ages 22--36.
 We explored T1 images, preprocessed with HCP-pipelines\footnote{https://github.com/Washington-University/HCPpipelines}.

For the morphometry data analysis we used Freesufrer \footnote{https://surfer.nmr.mgh.harvard.edu/} preprocessed features from section \textbf{Expanded FreeSurfer Data} for the same 1113 subjects. The morphometry characteristics as number of vertices, volumes, surface areas, and others were computed for $34$ cortical regions according to Desikan-Killiany Atlas and for $45$ subcortical areas according to Automated Subcortical Segmentation Atlas\cite{fischl2012freesurfer} summing up in $935$ vectorized features. 

\section{Methods}


\subsection{Morphometry data analysis and interpretation}

We used the morphometry data for gender classification with machine learning models in comparison as one of the most popular methods for data analysis in neuroimaging. The best performing model is chosen among different classifiers: XGBoost, k-Nearest Neighbors (KNN) and Logistic Regression (LR) with a grid-search. All considered models were validated with 10-fold cross-validation technique to give an understanding how the model will generalize to an independent dataset. 

\subsection{Full-size data analysis: 3D CNN}
 In this work we reproduced 3D CNN model architecture \cite{main_article} for the images sized \texttt{[58,70,58]}. The proposed network consists of three hidden layers, thus it is light, fast in training, and easy to interpret. To ensure the model stability, we performed 10-fold cross-validation with stratification to estimate model performance.

As the baseline for the 3D CNN network results we chose the support vector machine classifier with \texttt{rbf} kernel (SVM) on full sized images. This classifier is conventionally used as the network reference for image classification, for example on  \texttt{MNIST}. The SVM classifier from \texttt{sklearn} was trained on the full-size data reshaped to the 1-dimensional vector, as proposed in \cite{main_article}. We performed all 3D CNN experiments with \texttt{pytorch} framework, on Google Colab work station (Tesla K80 GPU).

\subsection{3D CNN interpretation}

\subsubsection{Meaningful Perturbations for 3D CNN results interpretation.}

The goal of the method \cite{meaningful_perturbation} is to perturb the smallest possible region of the MRI such that the model significantly changes its output probability for MR image class, which means that this region is the most important for model decision and it is the most informative part of the image. In this work we perturbed original image $x_0$ by replacing the corresponding region with Gaussian blurring of the image. Let $m:\Lambda \to [0,1]$ be a mask which map each voxel $u \in \Lambda$ of MRI to $m(u)$. Then the operation of perturbation of the MRI has the form: 
\begin{equation}
    P(x_0;m)=x_0\odot m+ (g_{\sigma_0 }\ast x_0)\odot(1-m),
\end{equation}
where $g_{\sigma_0}$ is a 3D Gaussian kernel with standard deviation $\sigma_0$. The closer $m(u)$ is to 1, the less perturbed the input image. The goal of the algorithm is to find the mask $m$ that leads to $f_c(P(x_0; m))\ll f_c(x_0)$, where $f_c(\cdot)$ is the probability of $x_0$ belonging to a class $c$, while perturbing the smallest part of the input image, i.e with a bigger amount of values $m(u)$ close to 1. To avoid the artifacts \cite{perturbation_artifacts}, we paded the $x_0$ with $j$ zeros and the mask $m$ applied to $x_0^K$: 
\begin{equation}
    x_0^K=x_0[K:H+j+K, K:W+j+K, K:D+j+K]
\end{equation}
 with integer $K$ drawn from the discrete uniform distribution on $[0, j )$, where $H,W,D$ --- size of the image. Also, we regularized $m$ in total-variation (TV) norm in low-resolution version, to force it had a more simple structure, upsampled it by factor $s$ to image size and applied a Gaussian filter on the full size mask. Let $M=g_{\sigma_m}\ast(Up(m,s))$, where $g_{\sigma_m}$ is a 3D Gaussian kernel with standard deviation $\sigma_m$. and $Up(\cdot ,s)$ is a trilinear upsampling algorithm  by factor of $s$. Finding $m_c$ for class $c$ can be formulated as the following optimization problem:
\begin{equation}
    m_c = \arg\min_{m} \mathbb{E}_{K\sim U[0,j)}\big[f_c(P(x_0^K,M))\big] +\lambda_2\sum_u\|\nabla M(u)\|^{\beta} + \lambda_1\|1-m \|_1.
\end{equation}
We perform ablation study and find following hyper-parameters which provide stable convergence: $\sigma_0=\sigma_m=10$, $\lambda_1=3$, $\lambda_2=1$, $\beta=7$, $s=4$, $j$ = 5, and we drawn $K$ from the discrete uniform distribution 10 times. The score is optimized by Adam optimizer with learning rate $\alpha = 0.3$, with exponential decay rates equals $\beta_1=0.9$, $\beta_2=0.99$. 

\subsubsection{Guided Backpropagation for 3D CNN results interpretation.}

In order to obtain saliency map for each voxel importance we used Guided Backpropagation method \cite{Guided_Backpropagation}. This approach computes the gradient of the score $y^c$ for class $c$ with respect to the input MRI image $x$ based on the network:
\begin{equation}
    \hat m_c = \frac{d y^c}{ d x}.
\end{equation}

The gradient is computed with specific backpropagation through the ReLU nonlinearity. In Guided Backpropagation, we backpropagate the positive values of
the gradient and set the negative ones to zero. 
Let $G^l$ be a gradient backpropagated through layer $l$ and $f_i^{l+1}=ReLU(f_i^l)$, then the used backpropagation process is given by the equation:
\begin{equation}
    G_{i}^{l} = (f_i^l>0)\cdot(G_i^{l+1}>0)\cdot G_i^{l+1}.
\end{equation}

After obtaining the saliency map in accordance with this specific backpropagation, 
we obtain an attention map $m_c = ReLU(\hat{m}_c)$ showing the voxels, which has only positive impact to score $y^c$.

\subsubsection{Grad CAM for 3D CNN results interpretation.}
Grad CAM \cite{GradCam} interprets the model, assuming that the deep CNN layers capture higher-level visual constructs \cite{chen2015microsoft}. In the neural network terminal layers we expected features maps to capture higher-level patterns which could be responsible for class discrimination. Grad CAM computes the gradient  for  score $y^c$  of  class $c$ before terminal layer  with respect to filter activations  of the last convolutional layer $F^k$. Then it computes the importance weights for each filter, i.e.
\begin{equation}
    \alpha_k^c = \frac{1}{H\cdot W\cdot D}\sum_{i=1}^H\sum_{j=1}^W\sum_{k=1}^D\frac{\partial y^c}{\partial F_{i,j,k}^k},
\end{equation}
where $H,W,D$ --- size of a filter activation tensor. $\alpha_k^c$ captures the ``importance'' of filter $k$ for a target class $c$. To obtain the class-discriminative
localization mask $m_c$, we computed a weighted combination of  filter activations, and follow it by a ReLU:
\begin{equation}
    m_c = ReLU\left[\sum_{k=1}\alpha_k^c \cdot F^k\right].
\end{equation}
Also we upsampled  $m_c$ to the input image resolution using trilinear interpolation.

\section{Results}

\subsection{Morphometry data}
\begin{figure}[ht]
  \centering
  \hspace*{-2cm}
  \includegraphics[scale=0.5]{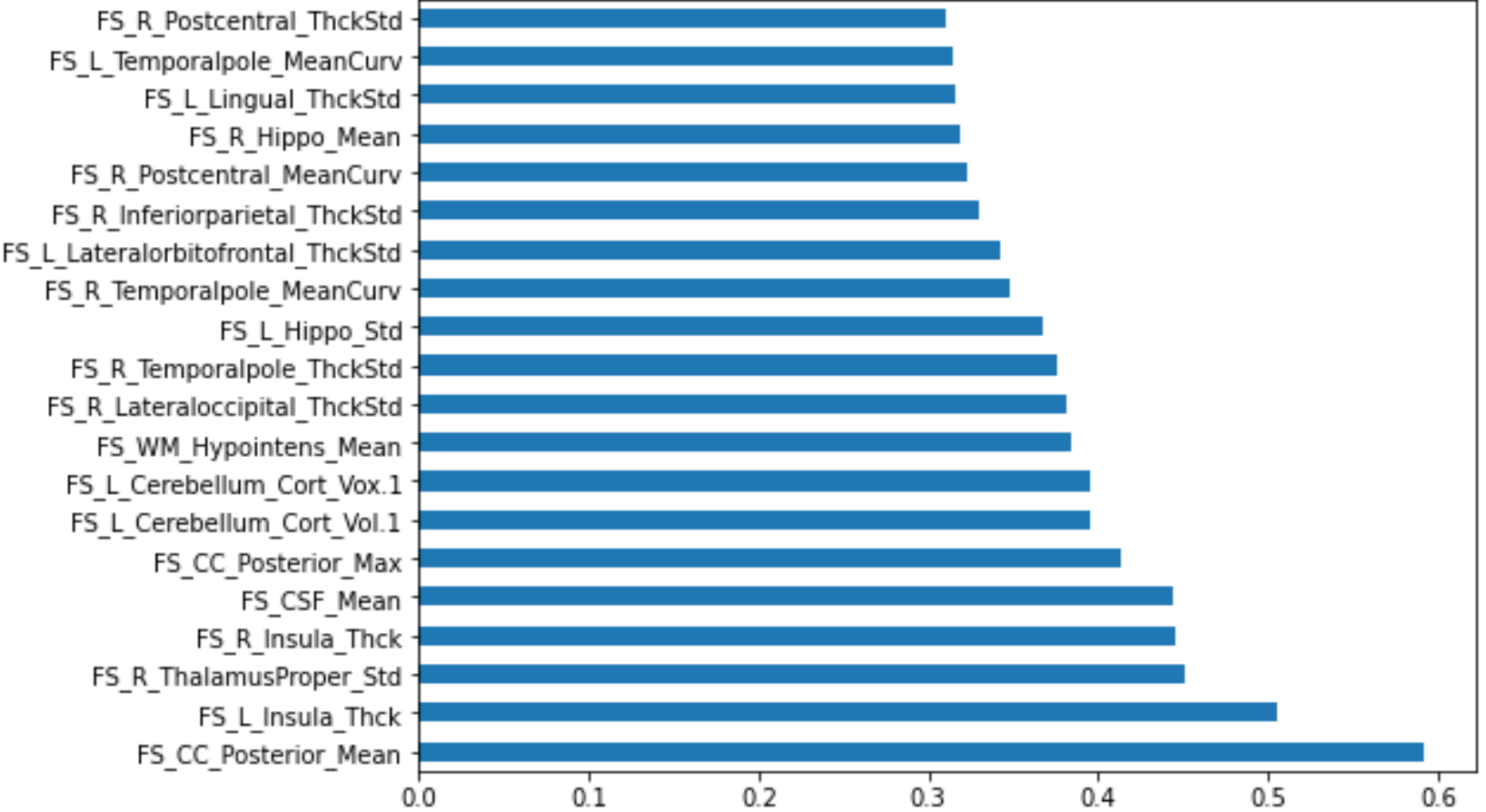}
  \caption{ Feature importances for Logistic Regression model.}
  \label{fig:feature_importances}
\end{figure}

The results of accuracy metrics for 10-folds cross-validation and 1113 subjects are in Table \ref{table:satellites} . Feature importances ($\beta$ scores) for Logistic Regression model chosen via a grid search are provided in Fig. \ref{fig:feature_importances}. Parameters of model are follow: C =1.3, penalty ='l2', max iter=10000. Importances were selected by coefficients of features in the decision function. Here on $x$ axis we provide the features scores and on $y$ axis --- names of features.

The classification models were chosen to explore the classification baseline accuracy across several different machine learning methods.

\begin{table}[ht]
  \caption{Results for baseline morphometry data classifcation models: 10-fold cross-validation}
  \centering
  \begin{tabular}{  p{100pt}  | p{75pt} | p{75pt} | p{75pt} }
 
  & XGB &  KNN  & LR  \\ \hline
  Mean accuracy  &   0.89 & 0.85 & 0.92   \\ \hline
  STD & 0.02 & 0.04 & 0.03 \\
 
  \hline
  \end{tabular}
  \label{table:satellites}
\end{table}

As can be seen from Fig. \ref{fig:feature_importances}, the most important features for the Logistic Regression model belong to the following brain regions volumes and intensities: corpus callosum, left and right insula and thalamic regions, as well as whole brian metrics for white matter hyperintensities. 

\subsection{3D CNN model results}

The 3D CNN model  yielded the mean accuracy of ${0.92 \pm 0.03}$ on 10-fold cross-validation.
SVM achieved ${0.90 \pm 0.02}$ accuracy on 10-fold cross-validation. So 3D-CNN model slightly outperformed standard SVM.

\subsection{Meaningful Perturbations for 3D CNN}
\begin{figure}[htp!]
\centering
\subfloat[\label{fig:meaningful perturbation}]{%
  \includegraphics[width=0.8\textwidth]{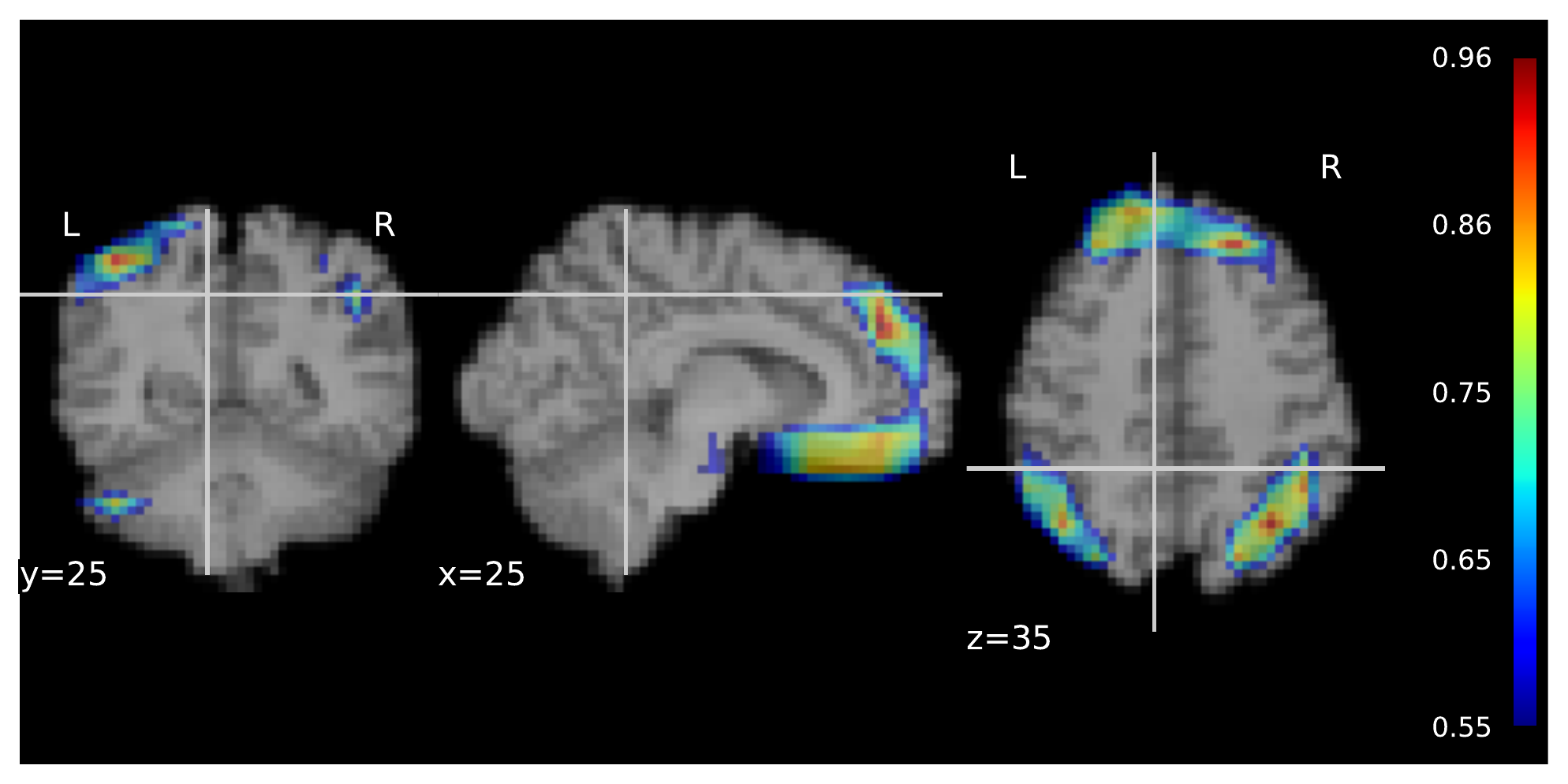}%
}\vfil
\subfloat[\label{fig:guided backpropagation}]{%
   \includegraphics[width=0.8\textwidth]{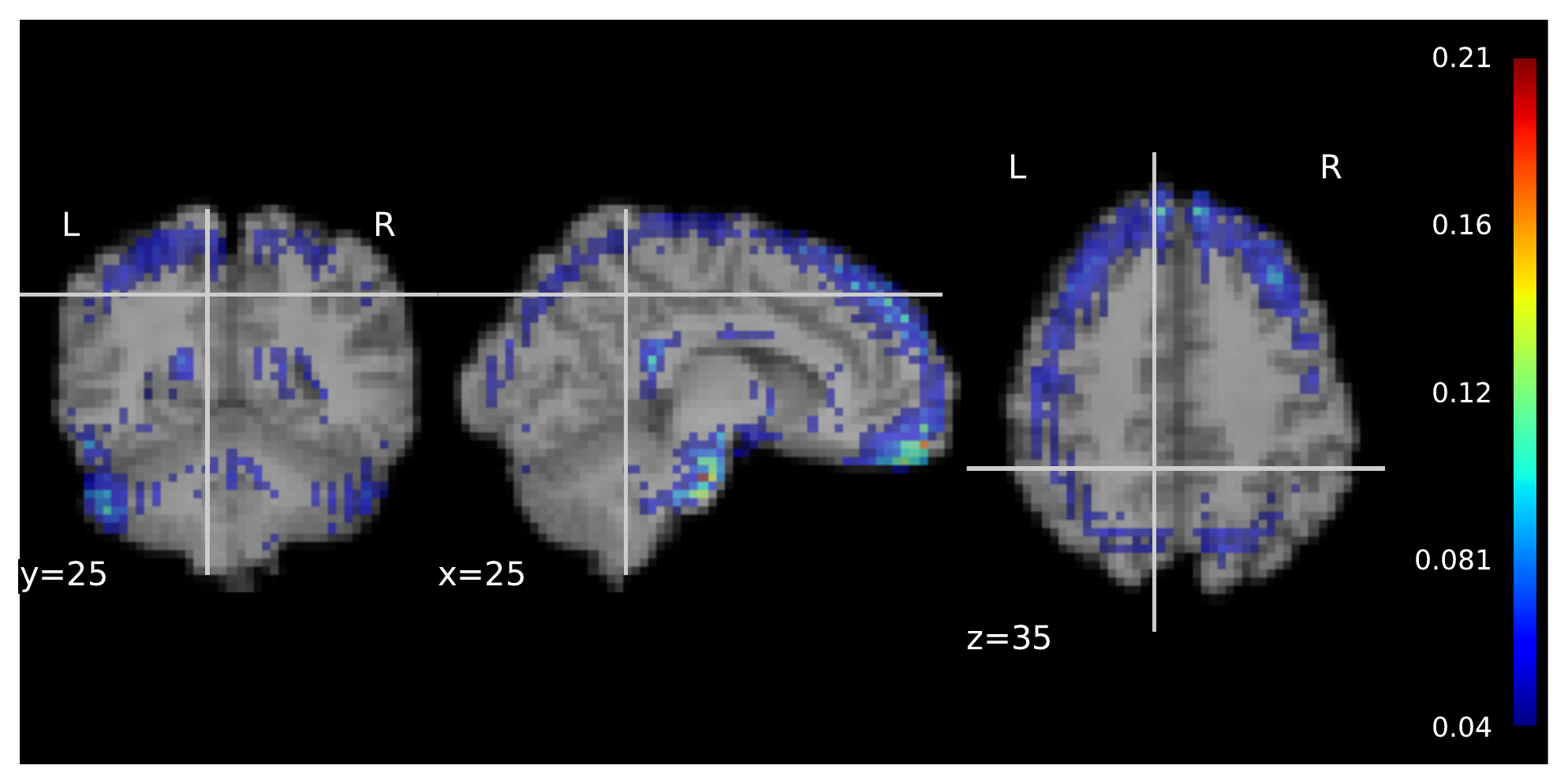}%
}\vfil
\subfloat[\label{fig:grad_cam}]{%
   \includegraphics[width=0.8\textwidth]{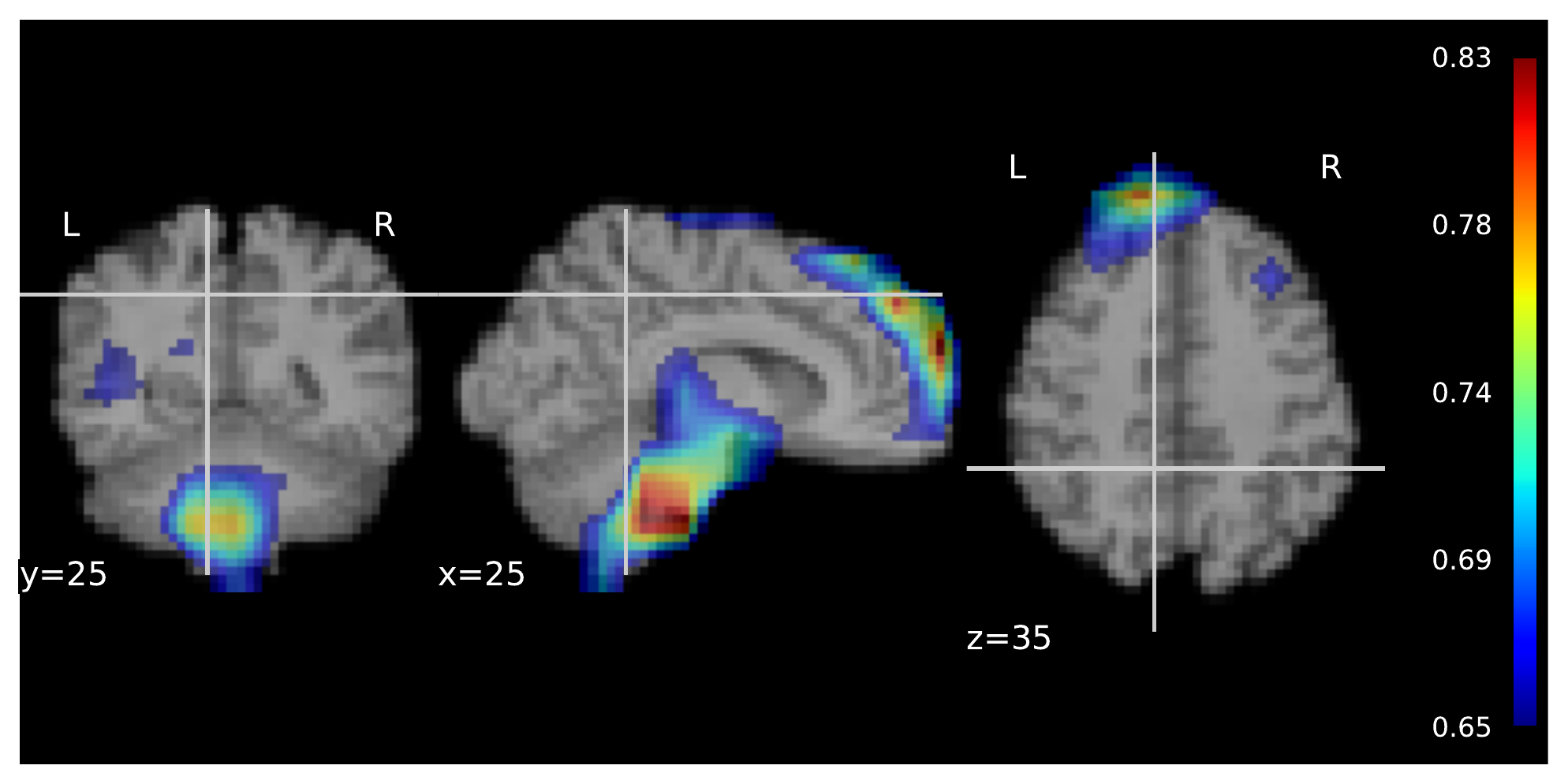}%
}
\caption[Three masks of the importance of brain regions obtained using]{Cross-sectional view on three attention maps for 3D CNN interpretation obtained with: a. Meaningful Perturbation (conjoined men and women attention mask), b. Guided Backpropagation, c. Grad CAM. The greater the voxel's value of each mask, the more important this voxel for classification.}
\end{figure}

For the two target classes in the gender differences classification, we got two different feature maps correspondingly with the meaningful perturbations algorithm. Two feature masks for men and women appeared to highlight different regions of interest and were then explored separately.


We completed 10 fold cross-validation to check the 3D CNN performance with the images restricted on masks. Then we multiplied every validation sample by averaging men mask, by averaging women mask or by the sum of these masks voxel-wise. The accuracy for the mask belonging to the men group is ${0.59 \pm 0.18}$, for the women mask --- ${0.59 \pm 0.07}$ and for the conjoined mask --- ${0.82 \pm 0.12}$  respectively. Thus, we can conclude that all necessary information for classification task is both in men and women masks (in their conjunction). The difference in masks for man and women may be explained by the specifics of the algorithm: we need to find the smallest region in the input image, deletion of which will decrease the probability of being a specific class. 
In Fig. \ref{fig:meaningful perturbation} we show the final mask which contains regions for men and women.

Next, to find the correspondence of the attention maps with specific brain regions we performed a segmentation of the MR images on 246 grey matter regions with the Human Brainnettome Atlas \cite{fan2016human}, as well as 50 white matter regions with ICBM-81 Atlas \cite{mori2006mri}. For each region of each brain atlas, we estimated fractions of voxels of this region included into the mask, which we have obtained via Meaningful Perturbations. We normalized these fractions, so  the values for all regions would sum up to 1. The top-5 scored regions of each atlas with the largest values are presented in Table \ref{tab:atlas}.

\begin{table}[ht]
\centering
\caption{The most discriminative regions of each atlas obtained with Meaningful Perturbations method.}
\footnotesize
\begin{tabular*}{\textwidth}{p{6.5cm}|p{2cm}}

\multicolumn{2}{c}{ICBM White-Matter Labels Atlas}\\ \hline
                                           Regions &  Score \\ \hline

                  Corticospinal tract right  &      0.1273 \\ \hline
Corticospinal tract left                        &        0.0927    \\ \hline
Anterior corona radiata right                     &            0.0594 \\ \hline
Pontine crossing tract                &   0.0580      \\\hline
Cerebral peduncle left          &                   0.0488    \\ \hline
\multicolumn{2}{c}{ }\\

\multicolumn{2}{c}{Human Brainnettome Atlas}\\ \hline
                                           Regions &  Score \\ \hline

 A13 Orbital gyrus Right    &   0.0131   \\ \hline
 A11L\_R, Orbital gyrus Right &    0.0124   \\ \hline
A39RD\_R, Inferior parietal lobule Right    &   0.0120  \\ \hline
A9\_46D\_R, Middle frontal gyrus Right    &   0.0118     \\ \hline
TI\_L, Parahippocampal gyrus Left     &   0.0117  \\ \hline

\end{tabular*}   
\label{tab:atlas}
\end{table}
\begin{tabular}{cc}
\end{tabular}

These findings partially overlap with the morphometry results, showing common white matter regions in the corpus callosum (Anterior corona radiata), as well as the cerebellum (Cerebral peduncle). The grey matter region in common overlaps on frontal gyri (Middle frontal gyrus Right).

\subsection{Guided Back-propagation for 3D CNN}
We computed a saliency map for every person in the dataset and then took mean over the dataset. As we have two classes in our dataset, the final map contains the regions of interest for each class, see Figure \ref{fig:guided backpropagation}.

\subsection{Grad CAM for 3D CNN}




We computed corresponding localization masks, containing information about both men and woman discriminative regions of interest. The cross-sectional view of the result is shown in Figure \ref{fig:grad_cam}.
To compare the resulting masks, we calculated the DICE\cite{milletari2016v} score between them.  For this, we chose a threshold value $t$ to obtain binary masks in the way to nullify all values outside the brain. The threshold of $t=0.65$ was chosen for Grad CAM and Meaningful Perturbations masks and $t=0.04$ for Guided Backpropagation. For Guided Backpropagation, such small threshold is used, due to the fact that the resulting mask is the gradient of the input image, and the exact values in the mask are much lower compared to other methods. The cross-sectional DICE score is shown in Table \ref{tab:compare}.
\begin{table}[ht]
\centering
\caption{DICE score between masks, MP - Meaningful Perturbations mask, GC - Grad CAM mask, GB - Guided Backpropagation mask}
\centering
  \begin{tabular}{  p{50pt}  | p{75pt} | p{75pt} | p{75pt} }
   & MP and GC & MP and GB & GC and GB \\
  \hline
  DICE  & 0.94  & 0.94 & 0.93  \\
  \hline
  \end{tabular}
\label{tab:compare}
\end{table}
\section{Discussion} 


 In the current work we aimed at studying gender-related differences in human brain by creating a predictive model to solve a classification task and to interpret its decisions.

In order to localize the most informative brain areas for classification task, we created attention maps for 3D CNN output in three different ways. Using these maps we were able to denote which brain regions play the most important role in gender classification. 
According to Meaningful  Perturbations method the brain regions with the highest classification accuracy  \ref{tab:atlas}  was Orbital Gyrus Right (A13\_R and A11l\_R) what corresponds with the results obtained from \cite{main_article}. Moreover we got that Inferior parietal lobule (A39RD\_R) is the region with high classification accuracy. That goes in line with the previous studies \cite{part_lobule}, where it was shown that parietal lobe activity is biased to the right hemisphere in men. We got that the  parahippocampal gyrus (TI\_L) plays part in gender differentiation. This result corresponds to the previous findings, where it was shown that parahippocampal gyrus activates more intensely in the female brain while executing large-scale spatial tasks \cite{yuan2019gender}. 
%



Right, and left corticospinal tracts showed high classification accuracy. This can be explained by the fact that the relative volume of the corticospinal tract on the left is larger in the female brain \cite{liu2011gender}. Anterior corona radiata right and cerebral peduncle are also responsible for gender differentiation what goes in line with \cite{main_article}. That confirms the most discriminative brain regions in morphometry model results, showing the importance of Corpus colosseum posterior part voxels mean (FS\_CC\_Posterior\_Mean) as the most discriminative in the Logistic Regression model. 

It is also worth noting, that attention maps in Fig. \ref{fig:meaningful perturbation} show the spatial pattern of frontoparietal resting-state brain network, which was initially discovered from resting-state fMRI activity and is thought to be involved in a wide variety of tasks by initiating and modulating cognitive control abilities \cite{cogcon}. It might be interesting in future research to look specifically at this network and explore it in terms of gender-related brain differences.


We show that the classification result on T1 images is lower than the result on FA images comparing the cross-validation mean scores (0.933), thus suggesting that DWI imaging might be more predictive for these type of structural differences. However, it should be mentioned that in the original approach as well, as in current paper, all T1 images were compressed to smaller tensors from \texttt{[260, 311, 260 ]} to \texttt{[58, 70, 58]} addressing memory issues. This compression may result in information loss and can explain similar classification score in neural networks comparing to the baseline methods.





\section{Conclusion} 

This work is an extension of the study on FA images from  \cite{main_article}. Here we create a 3D CNN model on T1 images of the same subjects and reveal similar gender-related differences in brain structure. The model exhibits the mean accuracy of $0.92 \pm 0.03$, which corresponds to the morphometry data classification and higher than SVM classification ($0.90 \pm 0.02$).

We apply several network interpretation methods to the 3D CNN model: Meaningful Perturbations, Grad CAM and Guided Backpropagation to find gender-specific patterns, and to compare their performances. We compare the results of interpretation in terms of the Dice coefficient. High scores on paired mask comparisons confirm that all three methods reveal similar patterns and thus belong to stable and trustworthy predictions. We found that Grad CAM and Guided Backpropagation are very sensitive to the slightest changes in the model weights, therefore, when using such layers as Dropout, it is necessary to fix the random seed. Then the Meaningful Perturbations method is less sensitive to small changes in model weights. We found that the Grad CAM method is the fastest one (executed within one minute) and ready plug-and-play method, Guided Backpropagation is slower (up to several minutes), as it propagates the gradient across the entire network, and has a more sparse attention map. The Meaningful Perturbations method is the slowest one (executes for several hours) yet showing the most anatomic-like attention maps. We recommend using the Grad CAM method for network interpretation at first in similar tasks.

Our deep learning model interpretation results are in line with the results from machine learning classification based on morphometry data. These findings are in line with previous studies as well, confirming the assumption that men's and women's differences more probably exist in the whole-brain range.

We also publish the code to the open-source library for public use. The proposed interpretation tool could be successfully used in various MRI pathology detection applications like epilepsy detection, Alzheimer's disorder diagnosis, Autism Spectrum disorder classification, and others.

\section{Acknowledgements} 
  The reported study was funded by RFBR according to the research project \textnumero{20-37-90149}.
  Also we acknowledge participation of Ruslan Rakhimov in development of meaningfull perturbation method on MRI data.

\bibliographystyle{splncs04}
\bibliography{main_paper}



                       


\newpage
\appendix
\section{The First Hidden Layer of 3D CNN Attention Analysis}
\label{appendinx_f}

We analyzed features obtained in First Hidden Layer of 3D CNN as in \cite{main_article}. Even though we used T1 modality MRI images in contorary to DWI modality and fractional anisotropy (FA) images in previous studies.

\begin{figure}[ht]
  \centering
  \begin{tabular}{ ll}
  \begin{tabular}{ccc}
    \includegraphics[width=.45\linewidth,height=150pt]{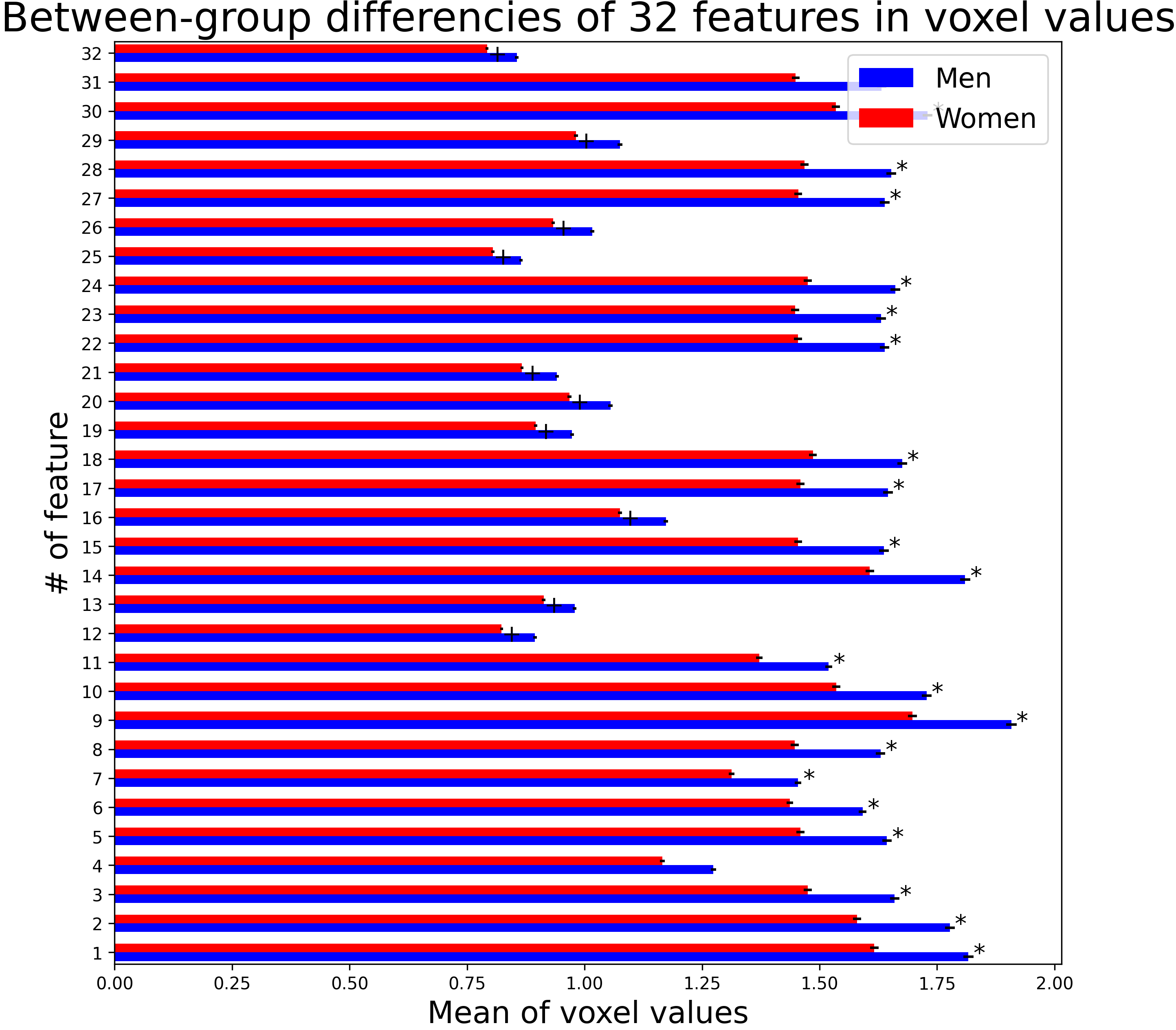} \\
    \small (a) 
  \end{tabular}
   \setlength{\tabcolsep}{10pt}
  \vspace{10pt}
  
  \begin{tabular}{ccc}
    \includegraphics[width=.45\linewidth,height=150pt]{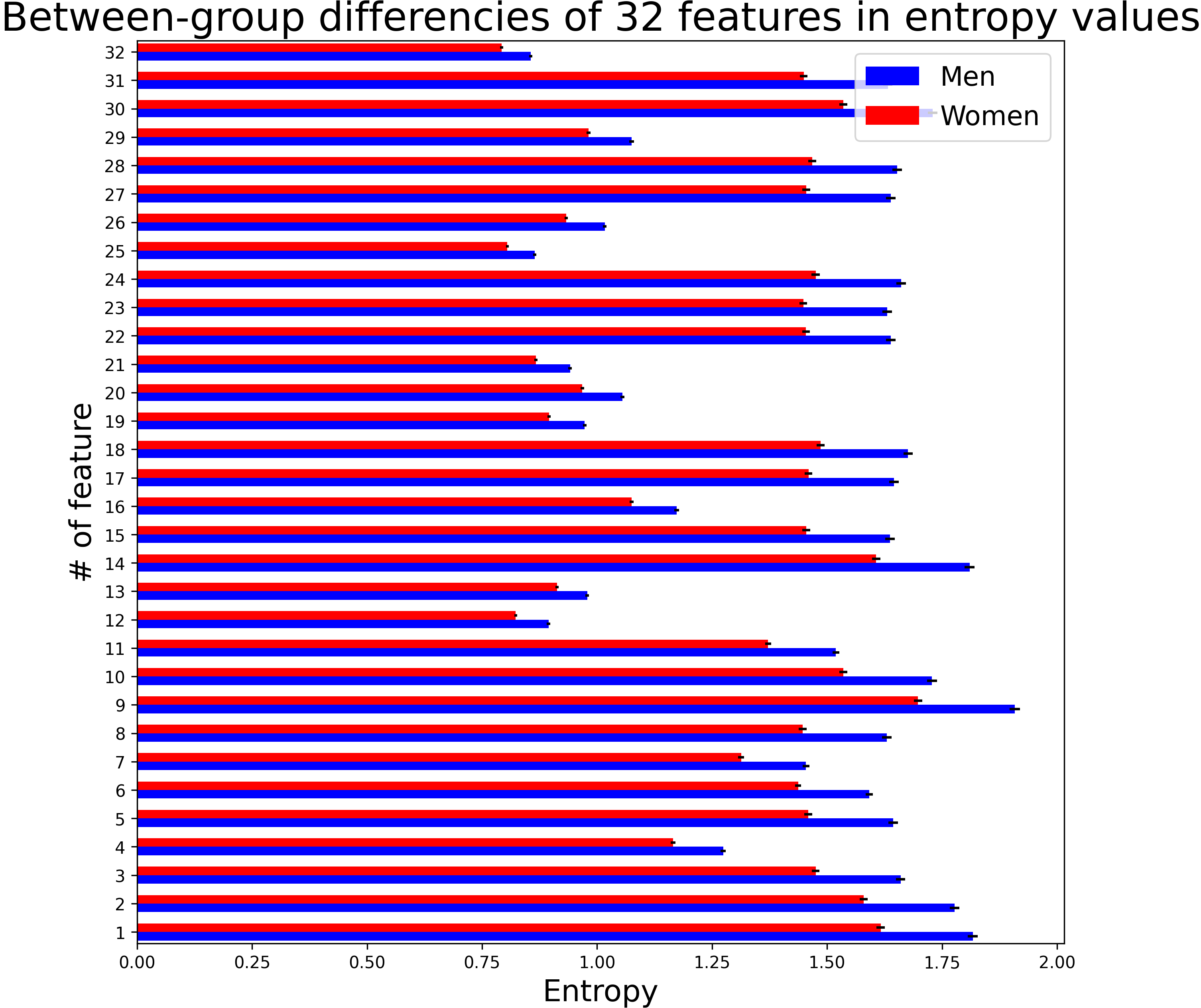} \\
    \small (b) 
  \end{tabular}
  \end{tabular}

  \caption{(a) Mean voxel values for each feature in men/women groups. Features that are significantly large for men are marked with *, features that are significantly large for women are marked with +. (b) Mean entropy values for each feature in men/women groups. }
    \label{fig:feature_mean}
\end{figure}

Similar to results shown on FA images, we found that mean voxel values for 31 features have a significant difference in men-women groups, with 10 features larger for women, and 21 features larger for men (see Fig.  \ref{fig:feature_mean}) accounting the multiple-comparisons correction. That reproduces the previously stated result, assuming that ``men's brains likely have more complex features as reflected by significantly higher entropy.'' As well as that important gender-related patterns are likely to be spread in the whole-brain grey and white matter. That highlights the importance of the results discussed in the main paper, as the attention maps compared from different approaches are extracted from the whole brain imagery, without any region-of-interest removal, as in \cite{main_article}.

\end{document}